\documentclass[12pt,epsf,amstex]{article}
\usepackage [dvips]{graphicx}
\usepackage{amsmath}
\usepackage{amssymb}
\usepackage{epsfig}

\addtocounter{secnumdepth}{1}
\begin{document}
\newcommand{\br}{\bar{r}}
\newcommand{\bbeta}{\bar{\beta}}
\newcommand{\bgamma}{\bar{\gamma}}
\newcommand{\tbeta}{\tilde{\beta}}
\newcommand{\tgamma}{\tilde{\gamma}}
\newcommand{\bE}{{\bf{E}}}
\newcommand{\bO}{{\bf{O}}}
\newcommand{\bR}{{\bf{R}}}
\newcommand{\bS}{{\bf{S}}}
\newcommand{\bT}{\mbox{\bf T}}
\newcommand{\bt}{\mbox{\bf t}}
\newcommand{\half}{\frac{1}{2}}
\newcommand{\summ}{\sum_{m=1}^n}
\newcommand{\sumq}{\sum_{q=1}^\infty}
\newcommand{\sumqno}{\sum_{q\neq 0}}
\newcommand{\prodm}{\prod_{m=1}^n}
\newcommand{\prodq}{\prod_{q=1}^\infty}
\newcommand{\maxm}{\max_{1\leq m\leq n}}
\newcommand{\maxphi}{\max_{0\leq\phi\leq 2\pi}}
\newcommand{\tsum}{\Sigma}
\newcommand{\bsA}{\mathbf{A}}
\newcommand{\bsV}{\mathbf{V}}
\newcommand{\bsE}{\mathbf{E}}
\newcommand{\bsT}{\mathbf{T}}
\newcommand{\bsZ}{\hat{\mathbf{Z}}}
\newcommand{\bse}{\mbox{\bf{1}}}
\newcommand{\bspsi}{\hat{\boldsymbol{\psi}}}
\newcommand{\cdottt}{\!\cdot\!}
\newcommand{\deltaR}{\delta\mspace{-1.5mu}R}
\newcommand{\invup}{\rule{0ex}{2ex}}

\newcommand{\bGamma}{\boldmath$\Gamma$\unboldmath}
\newcommand{\dd}{\mbox{d}}
\newcommand{\ee}{\mbox{e}}
\newcommand{\p}{\partial}
\newcommand{\expmVo}{\langle\ee^{-{\mathbb V}}\rangle_0}

\newcommand{\Rav}{R_{\rm av}}
\newcommand{\Rc}{R_{\rm c}}

\newcommand{\la}{\langle}
\newcommand{\ra}{\rangle}
\newcommand{\rao}{\rangle\raisebox{-.5ex}{$\!{}_0$}}  
\newcommand{\rae}{\rangle\raisebox{-.5ex}{$\!{}_1$}}
\newcommand{\raG}{\rangle_{_{\!G}}}
\newcommand{\rainr}{\rangle_r^{\rm in}}
\newcommand{\beq}{\begin{equation}}
\newcommand{\eeq}{\end{equation}}
\newcommand{\bea}{\begin{eqnarray}}
\newcommand{\eea}{\end{eqnarray}}
\def\lsim{\:\raisebox{-0.5ex}{$\stackrel{\textstyle<}{\sim}$}\:}
\def\gsim{\:\raisebox{-0.5ex}{$\stackrel{\textstyle>}{\sim}$}\:}

\title{\bf Central limit theorems for correlated variables: 
some critical remarks}

\author{{\bf H.J.~Hilhorst}\\[5mm]
{\small Laboratoire de Physique Th\'eorique, B\^atiment 210}\\[-1mm]
{\small Univ Paris-Sud XI and CNRS, 91405 Orsay, France}\\[-1mm]
{\small \texttt{Email: Henk.Hilhorst@th.u-psud.fr}}}

\maketitle

\begin{abstract}
In this talk I first review at an elementary level   
a selection of central limit theorems, including some lesser known cases,
for sums and maxima of uncorrelated and correlated random variables.
I recall why several of them appear in physics.
Next, I show that there is room for new versions of central limit theorems
applicable to specific classes of problems.
Finally, I argue that we have insufficient evidence 
that, as a consequence of such a theorem,
$q$-Gaussians occupy a special place in statistical physics.\\ 
\end{abstract}

\vspace{44mm}

\noindent {\small
Text at the basis of a talk presented at 
the 7th International Conference
on Nonextensive Statistical Mechanics: Foundations and Applications
(NEXT2008), Foz do Igua\c cu, Paran\'a, Brazil, 27-31 October 2008.}
\vspace{2mm}


\begin{flushright}{Preprint LPT Orsay 08/102}\\
\end{flushright}
\thispagestyle{empty}


Central limit theorems play an important role in physics, and in particular in
statistical physics. The reason is that this discipline deals almost always
with a very large number $N$ of variables, so that the limit $N\to\infty$
required in the mathematical limit theorems comes very close to being
realized in physical reality. 
Before looking at some hard questions, let us make an inventory 
of a few things we know.

\section{Sums of random variables}

\noindent {\bf Gaussians and why they occur in real life}\\

Let ${p}(x)$ be an arbitrary probability distribution 
of zero mean.
Draw from it independently $N$ variables $x_1, x_2,\ldots,x_N$
and then ask what is the probability $P_N(Y)$ that
the scaled sum $({x_1+x_2+\ldots+x_N})/N^{1/2}$ take the value $Y$.
The answer, as we explain to our students, is obtained by doing the
convolution 
$P_N(Y) = {p}(x_1)\ast{p}(x_2)\ast\cdots{p}(x_N)$.
After some elementary rewriting one gets
\beq
P_N(Y) = \frac{1}{\sqrt{2\pi{\la x^2\ra}}}\,
\exp\left(-\frac{Y^2}{2{\la x^2\ra}}\,+\,
\frac{1}{N^{1/2}}\big[\ldots\big]\right),
\label{derivationCLT}
\eeq
where the dots stand for an infinite series of terms that depend on all
moments of $p(x)$ higher than the second one,
$\la x^3\ra$, $\la x^4\ra$, \ldots. 
In the limit $N\to\infty$ the miracle occurs:
the dependence on these moments disappears from 
(\ref{derivationCLT}) and we find the Gaussian
$P_\infty(Y) = (2\pi{\la x^2\ra})^{1/2}\,
\exp\left(-\frac{1}{2}{Y^2}/\la x^2\ra\right)$.

The important point is that even if you didn't know beforehand about its
existence, this Gaussian results automatically
from any initially given $p(x)$ -- for example a binary distribution
with equal probability for $x=\pm 1$.
This is the Central Limit Theorem (CLT); it says that
the Gaussian is an {\it attractor\,} \cite{footnote0} {\it under addition\,} 
of independent identically distributed random variables.
An adapted version of the Central Limit Theorem
remains true for sufficiently weakly correlated variables.
\vspace{2mm}

This theorem of probability theory is, first of all, a mathematical truth.
In order to see why it is relevant to real life, 
we have to examine the equations of physics. It appears that these couple
their variables, in most cases, only over short distances and times,
so that the variables are effectively independent.
This is the principal reason for the ubiquitous
occurrence of Gaussians in physics (Brownian motion) and beyond
(coin tossing).
Inversely, the procedure of fitting a statistical curve by a Gaussian
may be considered to have a theoretical basis
if the quantity represented can be argued to arise 
from a large number of independent contributions, 
even if these cannot be explicitly identified.\\

\noindent {\bf L\'evy distributions}\\

{\it Symmetric L\'evy distributions.}
Obviously the calculation leading to (\ref{derivationCLT})
requires that the variance $\la x^2\ra$ be finite.
What if it isn't?
That happens, in particular, when for $x\to\pm\infty$
the distribution behaves as
$p(x)\simeq c_\pm |x|^{-1-\alpha}$ for some $\alpha\in(0,2)$.

Well, then there is a different central limit theorem.
The attractor is a L\'evy distribution $L_{\alpha,\beta}(Y)$,
where $Y$ is again the sum of the $x_n$ scaled with an appropriate 
power of $N$ and where $\beta\in(0,1)$ 
depends on the asymmetry between the amplitudes $c_+$ and $c_-$. 
A description of the $L_{\alpha,\beta}$ is given, {\it e.g.,} 
by Hughes (\cite{Hughes95}, see \S\,4.2-4.3).
In the symmetric case $c_+=c_-$ we have $\beta=0$.
Then $L_{1,0}(Y)$ is the Lorentz-Cauchy distribution
$P_\infty(Y)=1/[\pi(1+Y^2)]$ and $L_{2,0}(Y)$ is the Gaussian 
discussed above.\\

\begin{figure}
\begin{center}
\includegraphics[width=0.68\linewidth]{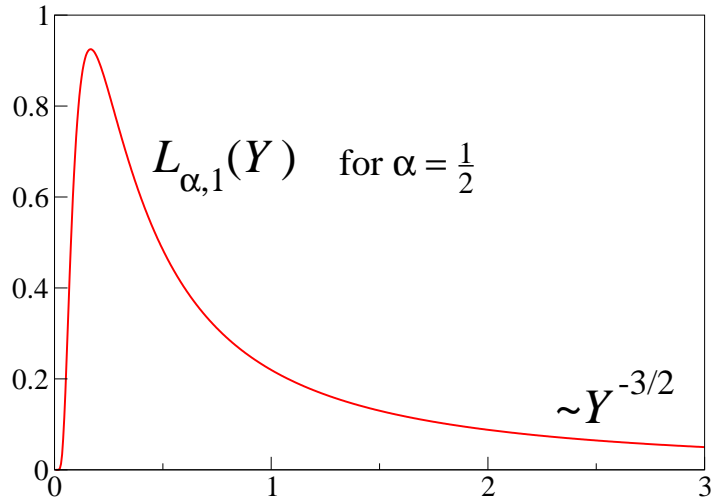}
\end{center}
\caption{\small The one-sided L\'evy distribution $L_{\frac{1}{2},1}(Y)$
is one possible outcome of a central limit theorem.}
\label{figlevyasb}
\end{figure}

{\it Asymmetric L\'evy distributions.}
If $c_-=0$, that is, if $p(x)$
has a slow power law decay only for large positive $x$, then
we have $\beta=1$ and 
the limit distribution is the {\it one-sided L\'evy distribution}.
In the special case $\alpha=\frac{1}{2}$,
shown in Fig.\,\ref{figlevyasb},
we have the Smirnov distribution $L_{\frac{1}{2},1}$.
It has the explicit analytic expression
\beq
P_\infty(Y)=\frac{1}{\sqrt{4\pi Y^3}}\exp\left(-\frac{1}{4Y}\right),
\qquad Y>0,
\label{Smirnovdistr}
\eeq
and decays for large $Y$ as $\sim Y^{-3/2}$.

All these L\'evy distributions are attractors 
under addition of random variables, just like the Gaussian,
and each has its own basin of attraction.\\

\noindent {\bf Addition of nonidentical variables}\\

\begin{figure}
\begin{center}
\includegraphics[width=0.6\linewidth]{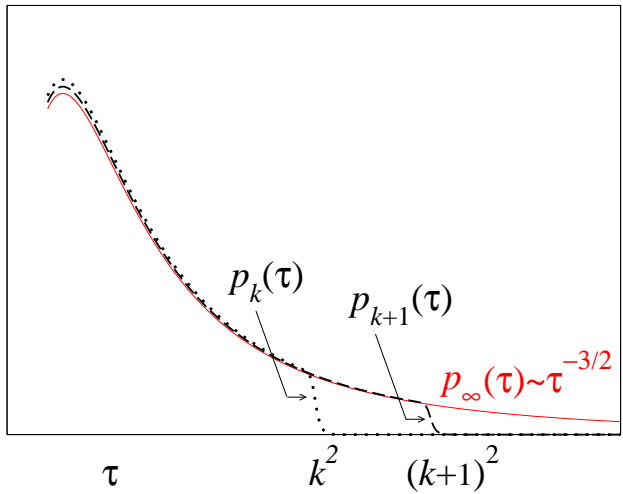}  
\end{center}
\caption{\small A sequence of probability distributions developing a ``fat
  tail'' proportional to $\sim \tau^{-3/2}$ when $k\to\infty$.}
\label{figptaunext}
\end{figure}

Mathematicians tell us that there do not exist any other attractors, 
at least not for sums of independent identically distributed (i.i.d.)
variables. However, suppose you add independent but {\it non-identical\,} 
variables.
If they're not too non-identical, you still get Gauss and L\'evy
distributions. The precise premises (the ``Lindeberg condition'')
under which the sum of a large number
of nonidentical variables is Gaussian distributed, 
may be found in a recent review for physicists
by Clusel and Bertin \cite{CluselBertin08}.

Now consider a case in which the Lindeberg condition does not hold.
Suppose a distribution $p_\infty(\tau)$, defined for $\tau>0$,
decays as $\tau^{-3/2}$ in the large-$\tau$ limit.
Let it be approximated,
as shown in Fig.\,\ref{figptaunext},
by a sequence of truncated distributions
$p_1(\tau), p_2(\tau),\ldots,p_k(\tau),\ldots$ 
which
is such that in the limit of large
$k$ the distribution $p_k(\tau)$ has its cutoff at $\tau\sim k^2$.
Then how will the sum $t_L\equiv\tau_1+\tau_2+\ldots+\tau_L$
be distributed?

The answer is that it will be a bell-shaped distribution which is 
neither Gaussian nor asymmetric L\'evy, but something in between. 
It's given by a complicated integral that I will not show here.
It is again an attractor: it does not depend on the shape of 
$p(\tau)$ and the $p_k(\tau)$ for finite $\tau$, but only on
the asymptotic large-$\tau$ behavior of these functions as well as on 
how the cutoff progresses for asymptotically large $k$.\\

\begin{figure}
\begin{center}
\includegraphics[width=0.6\linewidth]{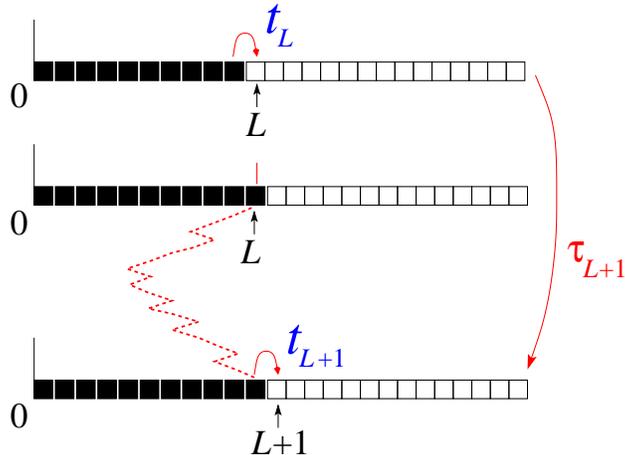}
\end{center}
\caption{\small A random walker on the positive 
half-line pays its first visit to site $L$ at time $t_L$. 
The sites (squares) already visited are colored black. 
The time interval between the first visits to $L$ and to $L+1$ is equal
to $\tau_{L+1}$. During this time interval the walker makes an excursion
in the direction of the origin, as indicated by the dotted trajectory.
The probability distributions $p_L(\tau_L)$ are independent 
but non-identical.}
\label{figwalknext}
\end{figure}

{\it An example: support of 1D simple random walk.}
An example of just these distributions occurs
in the following not totally unrealistic situation,
depicted in Fig.\,\ref{figwalknext}, and which
was studied by Hilhorst and Gomes \cite{HilhorstGomes98}.
A random walker on a one-dimensional lattice with reflecting boundary
conditions in the origin
visits site $L$ for the first time at time $t_L$. We can then write
$t_L=\tau_1+\tau_2+\ldots+\tau_L$, where $\tau_k$ is the time difference
between the first visit to the $(k-1)$th site and the $k$th site.
As $k$ increases, the $\tau_k$ tend to increase because of
longer and longer excursions inside the region already visited.
The $\tau_k$ are independent variables of the type described above.
For $L\to\infty$ the probability distribution of $t_L$
tends to an asymmetric bell-shaped function of the scaling 
variable $t_L/L^2$which is neither Gaussian nor L\'evy.
It is given by an integral that we will not present here.
It is universal in the sense that it depends only on
the asymptotic large $\tau$ behavior of the functions involved.\\

\noindent {\bf Sums having a random number of terms}\\

The game of summing variables still has other variations.
We may, for example, sum ${N}$ i.i.d. 
where ${N}$ itself is a random positive integer.
Let ${N}$ have a distribution $\pi_N(\nu)$ 
where $\nu$ is a continuous parameter 
such that $\la N\ra = \nu$. Then for $\nu\to\infty$ one easily derives
new variants of the Central Limit Theorem.\\

\section{Maxima of random variables}

\noindent {\bf Gumbel distributions}\\

Let us again start from $N$ independent identically distributed variables, 
but now ask a new question.
Let there be a given a probability law ${p}(x)$ which for large $x$
decays faster than any power law (it might be a Gaussian).
And suppose we draw $N$ independent random values $x_1,\ldots,x_N$ from this
law. We will set $Y=\max_{1\leq i\leq N}\,x_i$.
Then what is the probability distribution $P_N(Y)$ of $Y$ in the limit
$Y\to\infty$ ?  The expression is easily written down as an integral,
\beq
P_N(Y) \,=\,\frac{\dd}{\dd Y} \left(\,1\,-\,\int_{Y}^\infty\!\dd x\,
{p}(x)\right)^{\!N}. 
\label{exprmax}
\eeq
The calculation is a little harder to do than for the case of a sum. 
Let us subject the variable $Y$ to an appropriate (and generally
$N$-dependent) shift and scaling and again call the result $Y$. 
Then one obtains
\beq
P_\infty(Y)=\ee^{-Y}\,\ee^{-{\rm e}^{-Y}},
\label{Gumbel}
\eeq
which is the Gumbel distribution. 

The asymptotic decay of $p(x)$ was supposed here faster than any power law.
If it is as a power law, a different distribution appears,
called Fr\'echet;
and if $p(x)$ is strictly zero beyond some cutoff $x=x_{\rm c}$,
a third distribution appears, called Weibull.
Again, mathematicians tell us that
for this new question these three cases exhaust all possibilities.

In Ref.\,\cite{CluselBertin08} 
an interesting connection is established between
distributions of sums and of maxima.\\
 
\begin{figure}
\begin{center}
\includegraphics[width=0.7\linewidth]{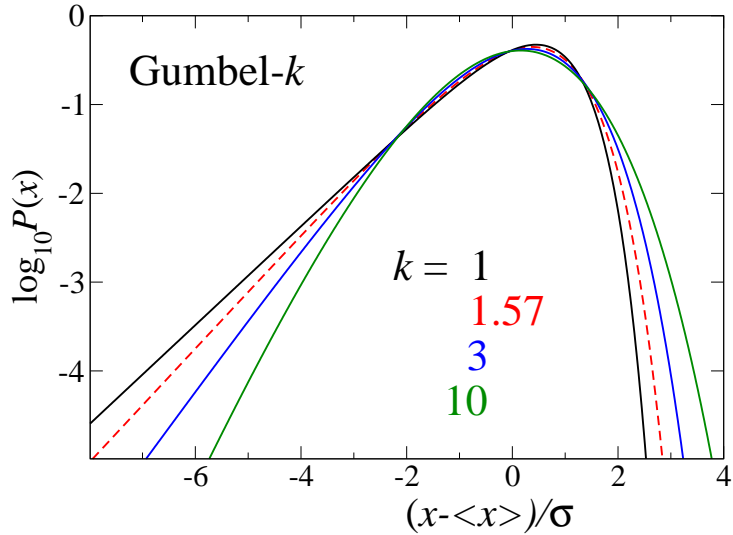}
\end{center}
\caption{\small The Gumbel-$k$ distribution for various values of $k$.}
\label{figgumbel}
\end{figure}

{\it The Gumbel-$k$ distribution.}
The Gumbel distribution (\ref{Gumbel}) is depicted in Figs.\,\ref{figgumbel},
where it is called ``Gumbel-1''.
This is because we may generalize the question and ask not how the largest
one of the $x_i$, but how the
$k$\,th largest one of them is distributed?
The answer is that it is a Gumbel distribution {\it of index\,} $k$.
Its analytic form is known and contains $k$ as a parameter.
For $k\to\infty$ 
it tends to a parabola, that is, to a Gaussian.

All these distributions are {\it attractors under the maximum operation}.
Even if you did not know them in advance,
you would be led to them starting from an arbitrary given distribution $p(x)$
within its basin of attraction.

Bertin and Clusel \cite{Bertin05,BertinClusel06} show that 
the definition of the Gumbel-$k$ distribution may be extended to real $k$.
These authors also show how Gumbel distributions of arbitrary index $k$
may be obtained as sums of correlated variables.
Their review article \cite{CluselBertin08} is particularly interesting.\\

\noindent{\bf The BHP distribution}\\

In 1998 Bramwell, Holdsworth, and Pinton (BHP) \cite{Bramwelletal98}
adopted a semi-empirical approach to the discovery of new universal
distributions.
These authors  noticed that, within error bars,
{\it exactly the same probability distribution} is observed for
(i) the experimentally measured power spectrum fluctuations of
3D turbulence; and (ii) the Monte Carlo simulated
magnetization of a 2D XY model
on an $L\times L$ lattice at temperature 
$T<T_{\rm c}$, in spin wave approximation.

\begin{figure}
\begin{center}
\includegraphics[width=0.7\linewidth]{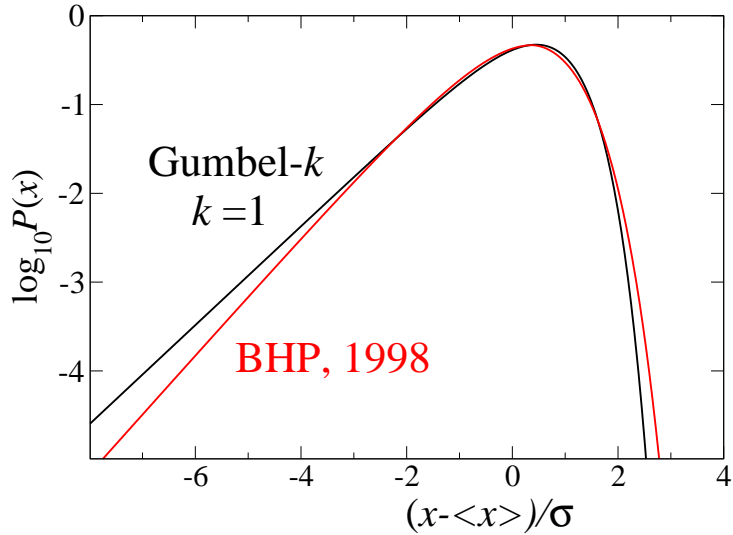}
\end{center}
\caption{\small The Gumbel-1 and the BHP distribution.}
\label{figgumbel7b}
\end{figure}

For the XY model Bramwell {\it et al.} \cite{Bramwelletal00,Bramwelletal01} 
were later able to calculate this distribution.  
It is given by a complicated integral that I will not
reproduce here and is called since the 
``BHP distribution.'' Fig.\,\ref{figgumbel7b} shows it
together with the Gumbel-1 distribution \cite{footnote1}.\\

{\it Numerical simulations.}
How universal exactly is the BHP distribution?
Bramwell {\it et al.} \cite{Bramwelletal00} were led to hypothesize that
the BHP occurs whenever you look for the maximum 
of, not independent, but {\it correlated\,} variables. 
To test this hypothesis these authors generated a random vector
$\vec{x}=(x_1,\ldots,x_N)$ of $N$ elements distributed independently according
to an exponential, and acted on it with a fixed random matrix
$\boldsymbol{M}$ such as to obtain $\vec{y}=\boldsymbol{M}\vec{x}$.
By varying $\vec{x}$ for a single fixed $\boldsymbol{M}$ they obtained the
distribution of $Y=\max_{1\leq i\leq N}\,y_i$
and concluded that indeed it was BHP.

However, Watkins {\it et al.} \cite{Watkinsetal02} showed one year later by an
analytic calculation that what appears to be a BHP distribution in reality 
crosses over to a Gumbel-1 law when $N$ is increased.
In this case, therefore, the correlation is irrelevant
and the attractor distribution is as for independent variables.

Watkins {\it et al.} conclude that ``even though subsequent results may show
that the BHP curve {\it can\,} result from strong correlation, it 
{\it need  not.}'' This example illustrates the danger of trying to attribute
an analytic expression to numerically obtained data.

In later work Clusel and Bertin \cite{CluselBertin08} present
heuristic arguments tending to explain why distributions 
closely resembling the BHP distribution occur so often in physics.\\

\noindent {\bf Wider occurrence of Gumbel and BHP?}\\

The Gumbel and BHP distribution have been advanced to fit curves
in situations where their occurrence is not {\it a priori\,} expected.
Two examples from the literature that appeared this month illustrate this.
Palassini \cite{Palassini08} performs Monte Carlo simulations that yield
the ground state energy of the Sherrington-Kirkpatrick model; 
this author fits his data 
by a\, Gumbel-6\, distribution (\cite{Palassini08}, Fig.\,4b).

Gon\c calves and Pinto \cite{GoncalvesPinto08} consider the distribution of the
cp daily return of two stock exchange indices (DJIA30 and S\&P100)
over a 21 year period. They find
that the cubic root of the square of this distribution is extremely well
fitted by the BHP curve (\cite{GoncalvesPinto08}, Figs.\,1 and 2).

In both cases the authors are right
to point out the quality of the fit.
But these examples also show that
having a very good fit doesn't mean you have a theoretical
explanation.\\ 

\section{Correlated variables}

In addition to the example discussed above, we will provide here two further
examples of how the maximum of a set of correlated random
variables may be distributed. 
These examples will illustrate the diversity of the results that emerge.\\

\begin{figure}
\begin{center}
\includegraphics[width=0.7\linewidth]{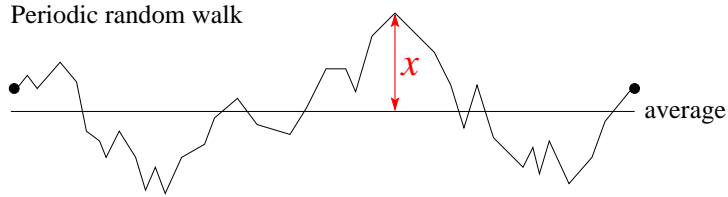}
\end{center}
\caption{\small A random walk on a finite interval with periodic boundary
  conditions. The random variable $x$ represents the maximum deviation of the
  walk from its interval average. The probability distribution $P(x)$ was
  calculated by Majumdar and Comtet \cite{MajumdarComtet04}.}
\label{rwmaxc}
\end{figure}

{\it Airy distribution.}
Fig.\,\ref{rwmaxc} shows the trajectory of a one-dimensional random walker
in a given time interval, subject to the condition that the starting point and
end point coincide. The walker's positions on two different times are clearly
correlated. Let $x$ denote the maximum deviation (in absolute value) of the
trajectory from its interval average.

Majumdar and Comtet \cite{MajumdarComtet04}
were able to show that this maximum distance is
described by the {\it Airy distribution\,}
(distinct from the well-known Airy function),
which is a weighted sum of hypergeometric functions that I will not reproduce
here. It is again universal: Schehr and Majumdar \cite{SchehrMajumdar06}
showed in analytic work, supported by numerical simulations,
that this same distribution appears for a wide class
of walks with short range steps. 
It turns out \cite{MajumdarComtet04}, however, 
that the distribution changes if the periodic boundary condition in time is
replaced by free boundaries. This therefore puts a limit on the universality
class \cite{footnote0}.\\

\begin{figure}
\begin{center}
\includegraphics[width=0.6\linewidth]{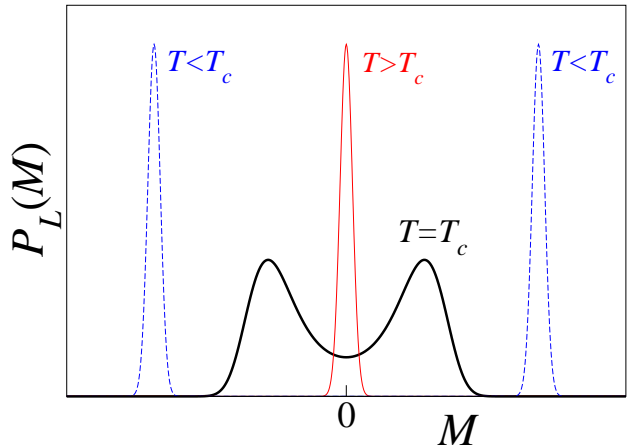}
\end{center}
\caption{\small Qualitative behavior of the
distribution $P_L(M)$ of the magnetization of the 2D Ising model on a
periodic $L\times L$ lattice. 
The sharp peaks for $T>T_{\rm c}$ and $T<T_{\rm c}$ are Gaussians.
For $T=T_{\rm c}$ and under suitable scaling $P_L(M)$ tends
in the limit $T\to\infty$ to a double-peaked universal 
distribution ${\cal P}(m)$; see Eq.\,(\ref{derivPM}) and the accompanying 
text.}
\label{figIsingnext}
\end{figure}

{\it Magnetization distribution of Ising 2D at criticality.}
We consider a finite $L\times L$ two-dimensional Ising model 
with a set of short-range interaction constants $\{J_k\}$.
Its magnetization (per spin) will be denoted $M=N^{-1}\sum_{i=1}^{N}s_i$, 
where $N=L^2$ and the $s_i$ are the individual spins. 
We ask what the distribution $P_L(M)$ is
exactly {\it at\,} the critical temperature $T=T_{\rm c}$.
This distribution can be determined, in principle at least, 
by a renormalization calculation
which in its final stage gives
\beq
L^{-\frac{1}{8}} P_L(M)= {\cal P}\left(m,\{L^{-y_\ell}u_\ell\}\right),
\label{derivPM}
\eeq
where $m=L^{\frac{1}{8}}M$ and where
$\{y_\ell\}$ is a set of positive fixed-point indices with corresponding
scaling fields $\{u_\ell\}$
({\it i.e.} the $u_\ell$ are nonlinear combinations of the $J_k$). 
In the limit $L\to\infty$ the dependence on these scaling fields disappears
and we have, in obvious notation, that
$\lim_{L\to\infty}L^{-\frac{1}{8}}P_L(M)={\cal P}(m)$.

In Fig.\,\ref{figIsingnext}
the distribution ${P}_L(M)$ is depicted qualitatively for $L\gg 1$
(it has two peaks!), together with the Gaussians that 
prevail when $T\neq T_{\rm c}$. The reason for $M$ not being
Gaussian distributed exactly at the critical point is that for 
$T=T_{\rm c}$ the spin pair correlation does not have an exponential but rather
a slow power law decay with distance: the spins are strongly correlated random
variables. 

The similarity between Eq.\,(\ref{derivPM}) and Eq.\,(\ref{derivationCLT})
is not fortuitous: the coarse-graining of the magnetization which is 
implicit in renormalization, amounts effectively to an addition of spin
variables; and the set of irrelevant scaling fields $\{u_\ell\}$ 
plays the same role as the set of higher moments 
$\{\la x^n\ra\,|\,n\geq 3\}$ in Eq.\,(\ref{derivationCLT}).

Eq.\,(\ref{derivPM}) says that 
${\cal P}(m)$ is an attractor under the renormalization group flow;
it is reached no matter what set of coupling constants $\{J_k\}$
was given at the outset. 
Here, too, there are limits on the basin of attraction: 
the shape of ${\cal P}(m)$ depends, in particular,
on the boundary conditions (periodic, free, or otherwise \cite{footnote3}).\\

The conclusion from everything above
is that attractor distributions come in all shapes 
and colors, and that it makes sense to try and discover new ones.

\section{$q$-Gaussians}

\begin{figure}
\begin{center}
\includegraphics[width=0.6\linewidth]{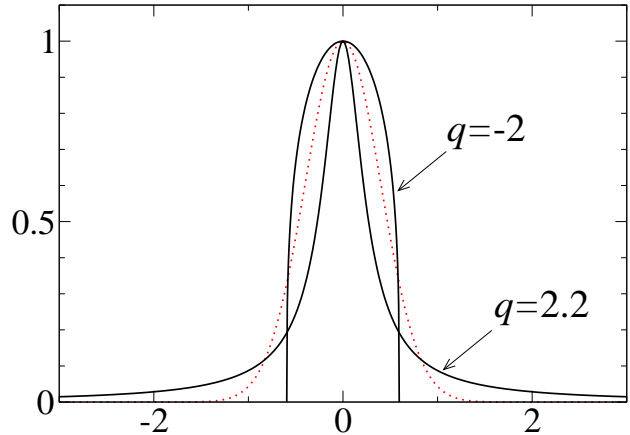}
\end{center}
\caption{\small Examples of $q$-Gaussians.
  Dotted curve: the ordinary Gaussian, $q=1$. 
  Solid curves: the
  $q$-Gaussians for $q=2.2$ and $q=-2$; the former has fat tails whereas the
  latter is confined to a compact support. All three curves are normalized to
  unity in the origin.}
\label{figqGnext}
\end{figure}

A {\it {$q$}-Gaussian\,} $G_q(x)$ is the power of a Lorentzian,
\beq
G_{{q}}(x)=\frac{\mbox{cst}}{\left[1+a x^2\right]^{\,p}}
= \frac{ \mbox{cst} }
{ \left[1+({q}-1)x^2\right]^{\frac{1}{{q}-1}} }\,,
\label{qGaussian}
\eeq
where in the second equality we have set $p=1/(q-1)$ 
and scaled $x$ such that $a=q-1$.
Examples of $q$-Gaussians are shown in Fig.\,\ref{figqGnext}. 
For ${q}=2$ the $q$-Gaussian is a Lorentzian;
in the limit ${q}\to 1$ it reduces to the ordinary Gaussian;
for ${q}<1$ it is a function with compact support,
defined only for $-x_{\rm m} <x< x_{\rm m}$ where 
$x_{\rm m}=1/\sqrt{1-q}$. For $q=0$ it is an arc of a parabola 
and for $q\to-\infty$ (with suitable rescaling of $x$)
it tends to a rectangular block.

Interest in $q$-Gaussians in connection with central limit theorems stems
from the fact that they have many remarkable properties that generalize
those of ordinary Gaussians.
One may consider, for example, the multivariate $q$-Gaussian 
obtained by replacing $x^2$ in
(\ref{qGaussian}) with $\sum_{\mu,\nu=1}^n x_\mu A_{\mu\nu} x_\nu$
(with $A$ a symmetric positive definite matrix).
Upon integrating this $q$-Gaussian on $m$ of its variables we find that
the marginal $(n-m)$-variable distribution is $q_m$-Gaussian 
with $q_m=1-2(1-q)/[2+m(1-q)]$
(see Vignat and Plastino \cite{VignatPlastino07}; this relation seems to have 
first appeared in Mendes and Tsallis \cite{MendesTsallis01}). 

A special case is the uniform probability distribution inside
an $n$-dimensional sphere of radius $R$, 
\beq
P_n(x_1,\ldots,x_n) = \mbox{cst}\times
\theta \left( R^2-\sum_{\mu=1}^n x_\mu^2 \right),
\label{unifsphere}
\eeq
where $\theta$ denotes the Heaviside step function.
This is actually a multivariate $q$-Gaussian with $q=-\infty$.
Integrating on $m$ of its variables yields a $q$-Gaussian with $q_m=1-2/m$.
We see that for large $m$ both in the general and in the special case 
$q_m$ approaches unity and hence these marginal distributions tend under
iterated tracing to an ordinary Gaussian shape.

Let us first see, now, how $q$-Gaussians may arise as solutions of certain
partial differential equations in physics.\\

\noindent{\bf Differential equations and $q$-Gaussians}\\

{\it Thermal diffusion in a potential.}
The standard Fokker-Planck (FP) equation describing a particle of 
coordinate $x$ diffusing at a temperature $T$ in a potential $U(x)$ reads
\beq
\frac{\partial P(x,t)}{\partial t} = \frac{\partial}{\partial x}
\left[ {U'(x)}\, P \,+\, 
k_BT\,\frac{\partial P}{\partial x}\, \right].
\label{FP}
\eeq
Its stationary distribution $P_U^{\rm st}(x)$
is the Boltzmann equilibrium in that potential,
$P_U^{\rm st}(x) = \mbox{cst}\times\exp[-\beta U(x)]$,
where $\beta=1/k_BT$.
For the special choice of potential
$U'(x)=\alpha x/(1+\gamma x^2)$
the stationary distribution becomes the $q$-Gaussian
\beq
P_U^{\rm st}(x) = { \mbox{cst}\times
\left(1+\gamma x^2\right)^{-{\alpha\beta}/{\gamma}}} .
\label{solqG}
\eeq
This distribution is an {\it attractor under time evolution,} 
the latter being defined by the
FP equation (\ref{FP}); a large class of reasonable initial
distributions will tend to (\ref{solqG}) as $t\to\infty$
\cite{footnote4}. It should be noted, however, that 
by adjusting $U(x)$ we may 
obtain any desired stationary distribution,
and hence the $q$-Gaussian of Eq.\,(\ref{solqG}) plays no exceptional role.

The following observation is trivial but will be of interest
later on in this talk.
Let $x(t)$ be the Brownian trajectory of the diffusing particle.
Let $x(0)$ be arbitrary and let
$x(t)$, for $t>0$, be the stochastic solution of
the Langevin equation associated  \cite{VanKampen92} with the
FP equation (\ref{FP}).
Let $\xi_n=x(n\tau)-x((n-1)\tau)$, where $\tau$ is a finite time interval.
Then $Y_N=\xi_1+\xi_2+\ldots+\xi_N$ (without any scaling)
is a sum which for $N\to\infty$ has the distribution $P_U^{\rm st}(Y)$.
In particular, if $U(x)$ is chosen such as to yield (\ref{solqG}),
we have constructed a $q$-Gaussian distributed sum.\\

{\it Finite difference scheme} \cite{footnote7}. 
Rodr\'{\i}guez {\it et al.\,}\cite{Rodriguezetal08} recently studied
the linear finite difference scheme 
\beq
r_{N,n}+r_{N,n+1}=r_{N-1,n}\,,
\label{finitediff}
\eeq
where $N=0,1,2,\ldots$ and $n=0,1,\ldots,N$. 
The quantity $p_{N,n}\equiv\binom{N}{n}r_{N,n}$ may be interpreted
as the probability that a sum 
of $N$ identical correlated binary variables be equal to $n$.
For specific boundary conditions,
the authors were quite remarkably able to find a class
of analytic solutions to Eq.\,(\ref{finitediff}) and
observed that the $N\to\infty$ limit of the sum law $p_{N,n}$
is a $q$-Gaussian.

To understand better what is happening here,
let us set $t=1/N$, $x=1-2n/N$, and $P(x,t) = N\binom{N}{n}r_{N,n}$.
When expanding Eq.\,(\ref{finitediff}) in powers of $N^{-1}$ 
one discovers \cite{Hilhorst09} that
$P(x,t)$ satisfies the Fokker-Planck equation
\beq
\frac{\partial P(x,t)}{\partial t} = \tfrac{1}{2}
\frac{\partial^2}{\partial x^2}\left[(1-x^2)P\right]
\label{FPfinitediff}
\eeq
for $t>0$ and $-1<x<1$.
The ``time'' $t$ runs in the direction of {\it decreasing\,} $N$.
Hence Rodr\'{\i}guez {\it et al.}
have solved a parabolic equation backward in time
and determined, starting from the small-$N$ behavior, 
what is actually an {\it initial\,} condition at $N=\infty$.
It is obvious that $q$-Gaussians are not singled out here:
there exists a solution to Eq.\,(\ref{FPfinitediff})  
for any other initial condition at $t=0$, 
and concomitantly to Eq.\,(\ref{finitediff}) 
for any desired limit function $p_{\infty,n}$ at $N=\infty$.

Therefore, in this and the preceding paragraph, the occurrence of
$q$-Gaussians in connection with Fokker-Planck equations cannot be
construed as an indication of a new central limit theorem.\\

{\it The porous medium equation.}
Let us consider a fluid flowing through a porous medium.
Three equations of physics provide the basic input
for the description of this flow, namely
(i) the continuity equation for the fluid density $\rho(\vec{x},t)$;
(ii) Darcy's law, which relates the fluid velocity $\vec{v}$ to its pressure
$p$ by $\vec{v}=-\mbox{cst}\times\vec{\nabla} p$;
and (iii) the adiabatic equation of state of the ideal gas.
Upon combining these one finds the
{\it porous medium equation}
\beq
{\frac{\partial\rho}{\partial t} = \Delta \rho^{2-q}}, 
\qquad  q=1-C_p/C_v\,,
\label{pormedu}
\eeq
where $C_p/C_v$ is the specific heat ratio. 
For $q=1$ this equation reduces to the ordinary diffusion equation. 

Equation (\ref{pormedu}) is nonlinear and its general solution, 
{\it i.e.,} for an arbitrary initial condition $u(\vec{x},t)$, 
cannot be found.
It is however possible to find special classes of solutions.
One special solution is obtained by looking for solutions that are
(i) radially symmetric, {\it i.e.,} dependent only on $x\equiv|\vec{x}|$;
and (ii) scale as $u(\vec{x},t)=t^{-db} F(xt^{-b})$. 
After scaling of $x$ and $t$
we obtain the similarity solution
\beq
\rho(\vec{x},t)= \frac{c_0}{t^{db}}
\left[\,1\,+\,(q-1)\,\frac{x^2}{t^{2b}}\,\right]^{-\frac{1}{q-1}},
\label{pormedsol}
\eeq
in which $b = 1/[d(1-q)+2]$ and where also
$c_0$ is uniquely defined in terms of the parameters of the equation.
Mathematicians (see {\it e.g.}\,\cite{Vazquez06})
have shown that initial distributions 
with compact support tend asymptotically
towards this similarity solution.
The asymptotic behavior (\ref{pormedsol}) is
conceivably robust, within a certain range, 
against various perturbations of the porous medium equation.
It is not clear to me if and how this property can be connected 
to a central limit theorem.\\

\noindent {\bf $q$-statistical mechanics}\\ 

Considerations from a $q$-generalized statistical mechanics 
\cite{Tsallis88,GellMannTsallis04,Tsallis09} have led
Tsallis \cite{Tsallis05} to surmise that in the limit $N\to\infty$ 
the sum of $N$ correlated random variables becomes,
under appropriate conditions, $q$-Gaussian distributed;
that is, on this hypothesis $q$-Gaussians are attractors 
in a similar sense as ordinary Gaussians.
Now, variables can be correlated in very many ways. 
To fully describe $N$ correlated random variables you need
the $N$ variable distribution {$P_N(x_1,\ldots,x_N)$}. 
Taking the limit $N\to\infty$ requires knowing the {\it set of functions}
\beq
P_N(x_1,\ldots,x_N),  \qquad N=1,2,3,\ldots
\eeq
In physical systems 
the $P_N$ are determined by the laws of nature;
the relative spatial and/or temporal coordinates of the variables,
usually play an essential role. 
The examples of the Ising model and of the Airy distribution show
how widely the probability distributions of
strongly correlated variables may vary.
Hence, in the absence of any
elements of knowledge about the physical system that they describe,
statements of uniform validity about correlated variables 
cannot be expected to be very specific.\\

\noindent {\bf $q$-Central Limit Theorem}\\

We now turn to a $q$-generalized central limit theorem ($q$-CLT) 
formulated by Umarov {\it et al.} \cite{Umarovetal08}.
It says, essentially, the following.
Given an infinite set of random variables $x_1, x_2, \ldots, x_n,\ldots,$ 
let the first $N$ of them be correlated according to a certain 
condition {${\cal C}_N(q)$}, where $N=1,2,3,\ldots$.
Then the partial sum $Y_N=\sum_{n=1}^Nx_n$, after appropriate scaling 
and in the limit $N\to\infty$, 
is distributed according to a $q$-Gaussian.
The theorem is restricted to $1<q<2$.  
The conditions ${\cal C}_N(q)$ are concisely referred to as 
``$q$-independence'' in Ref.\,\cite{Umarovetal08} and
for $q=1$ reduce to the usual condition of random variables being
independent. 
Closer inspection of the theorem prompts two questions.

First, the conditions ${\cal C}_N(q)$
are difficult to handle analytically.
If a theoretical model is defined by means of its 
$P_N(x_1,\ldots,x_N)$ for $N=1,2,3,\ldots$,
then one would have to check that these satisfy the ${\cal C}_N(q)$.
I am not aware of cases for which this has been possible.
In the absence of examples it is hard to see
why nature would generate exactly this type of
correlations among its variables.

Secondly, the proof of the theorem makes use of ``$q$-Fourier space,''
the $q$-Fourier transform ($q$-FT) having been defined in 
Ref.\,\cite{Umarovetal08} as a generalization of the ordinary FT. 
The $q$-FT has the feature that when applied to a $q$-Gaussian 
it yields a $q'$-Gaussian with $q'=(1+q)/(3-q)$, for $1\leq q<3$.
Now the $q$-FT is a nonlinear mapping which appears not to have an inverse
\cite{footnote2}. It is therefore unclear at present how the statements 
of the theorem derived in $q$-Fourier space can be translated 
back in a unique way to ``real'' space.

\section{The search for $q$-Gaussians}

\noindent {\bf Mean-field models}\\

\begin{figure}
\begin{center}
\includegraphics[width=0.6\linewidth]{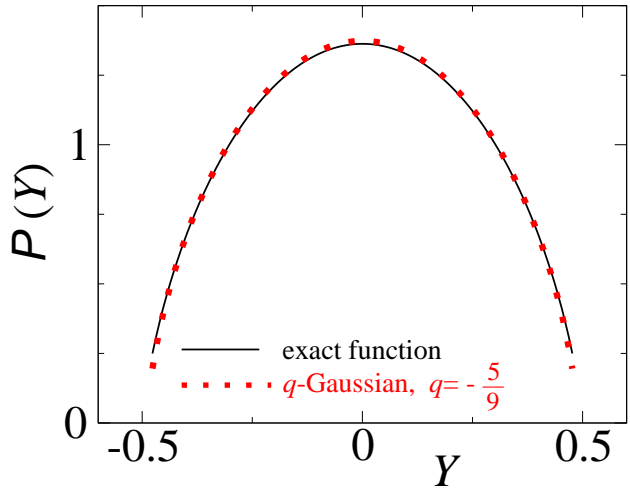}
\end{center}
\caption{\small
Comparison of the $q$-Gaussian $G_q(Y)$ (dotted curve)
guessed in Ref.\,\cite{Thistletonetal06}
on the basis of numerical data and
the exact distribution $P(Y)$ (Eq.\,(\ref{PHS}), solid curce) calculated in 
Ref.\,\cite{HilhorstSchehr07}.
The curves are for $\rho=\frac{7}{10}$ and
the $q$-Gaussian has $q=-\frac{5}{9}$.
The difference between the two curves is of the order of the
thickness of the lines and just barely visible to the eye.}
\label{figPUnextb}
\end{figure}

Independently of this $q$-CLT Thistleton {\it et al.} \cite{Thistletonetal06}
(see also Ref.\,\cite{Moyanoetal06}) 
attempted to see a $q$-Gaussian arise in a numerical experiment.
These authors defined a system of $N$ variables $x_i$, $i=1,2,\ldots,N$, 
equivalent under permutation. Each variable is drawn from a uniform 
distribution on the interval $(-\frac{1}{2},\frac{1}{2})$ but 
the $x_i$ are correlated in such a way that 
$\la x_jx_k\ra=\rho\la x_1^2\ra$ for all $j\neq k$,
where $\rho$ is a parameter in $(0,1)$ \cite{footnote5}.
They considered the sum $Y_N=(x_1+\ldots+x_N)/N$ and determined
its distribution $P(Y)$ in the limit $N\gg 1$.  
For $\rho=\frac{7}{10}$ the numerical results for $P(Y)$
can be fitted very well by a $q$-Gaussian $G_q(Y)$
with $q=-\frac{5}{9}$, shown as the dotted curve in Fig.\,\ref{figPUnextb}.
This system of correlated variables is sufficiently simple that Hilhorst and
Schehr \cite{HilhorstSchehr07} were able to do the analytic calculation of the
distribution. They found that $Y$ is distributed according to
\beq
P(Y)= \Big( \frac{2-\rho}{\rho} \Big)^\half
         \exp \Big(\! -\frac{2(1-\rho)}{\rho} 
                      \big[ \,\mbox{erf}^{-1}(2Y)\, \big]^2\, \Big),         
\label{PHS}
\eeq
for $-\tfrac{1}{2}<Y<\tfrac{1}{2}$ shown as the solid curve in
Fig.\,\ref{figPUnextb}.
The difference between the exact curve and the $q$-Gaussian
approximation is of the order of the thickness of the lines.
More importantly, the calculation of Ref.\,\cite{HilhorstSchehr07} 
shows that the distribution of the sum $Y$
varies with the initially given one of the $x_i$.
This initial distribution may be
fine-tuned such as to lead for $N\to\infty$
to almost any limit function $P(Y)$ -- in particular, to a
$q$-Gaussian. The existence of $q$-Gaussian distributed sums was already
pointed out below Eq.\,(\ref{solqG}) and is no surprise. However,
there is, here no more than in the case of the FP equation, 
any indication that distinguishes $q$-Gaussians from other functions. 

The work discussed here concern a mean-field type model: there is
full permutational symmetry between all variables. This will be different in
the last two models that we will now take a look at.\\

\noindent {\bf Logistic map and HMF model}\\

Two well-known models of statistical physics have been evoked 
several times by participants \cite{Tirnakli08,Rapisarda08} at this meeting.
The common feature is that in each of them the variable studied is obtained as
an average along a deterministic trajectory.\\

{\it Logistic map.} 
In their search for occurrences of $q$-Gaussians in nature, 
Tirnakli {\it et al.} \cite{Tirnaklietal07}
considered the logistic map
\beq
x_{\ell}=a-x^2_{\ell-1}\,, \qquad \ell=1,2,\ldots
\label{logisticmap}
\eeq
A motivation for this choice is the appearance 
\cite{BaldovinRobledo02} of $q$-{\it exponentials\,} 
in the study of this map.
Starting from a uniformly random initial condition $x=x_0$,
%
Tirnakli {\it et al.}
determined the probability distribution of the sum 
\beq
Y=\sum_{\ell=n_0}^{n_0+N} x_\ell
\label{defy}
\eeq
of successive iterates, scaled with an appropriate power of $N$,
in the limit $N\gg 1$. 
Their initial report of $q$-Gaussian behavior
at the Feigenbaum critical point (defined by a critical value
$a=a_{\rm c}$) was critized by Grassberger \cite{Grassberger08}.
Inspired by a detailed study due to Robledo and Mayano
\cite{RobledoMayano08}, who connect
properties {\it at\,} $a=a_{\rm c}$ to properties observed on {\it
approaching\,} this critical point, 
Tirnakli {\it et al.} \cite{Tirnaklietal08} 
took a renewed look at the same question and now see indications for a
$q$-Gaussian distribution of $Y$ {\it near\,} the critical point.\\

{\it Hamiltonian Mean Field Model.}
The Hamiltonian mean-field model (HMF), introduced in 1995 by
Antoni and Ruffo \cite{AntoniRuffo95}, describes $L$ unit
masses that move on a circle subject to a mean field potential. 
The Hamiltonian is, explicitly,
\beq
{\cal H} = \sum_i\frac{p_i^2}{2} \,+\, \frac{1}{2L}\sum_{i,j}
\left[ 1-\cos(\theta_i-\theta_j) \right],
\label{HMFmodel}
\eeq 
where $p_i$ and $\theta_i$ are the momentum and the polar angle, respectively,
of the $i$th mass. The angles were originally considered to describe the state
of classical XY spins, so that $\vec{m}_i=(\cos\theta_i,\sin\theta_i)$
is the magnetization of the $i$th spin.
The HMF has a solvable equilibrium state. 
At a critical value $U=U_{\rm c}=0.75$ of the total energy per particle
a phase transition occurs from a high-temperature state with uniformly
distributed particles to a low-temperature one with a spontaneous
value of the ``magnetization'' $\la|\vec{M}|\ra$, where 
$\vec{M}=L^{-1}\sum_{i=1}^L\vec{m}_i$.
When launched with certain nonequilibrium initial conditions,
the system, before relaxing to equilibrium, appears to enter 
a ``quasi-stationary state'' (QSS) whose lifetime diverges with $N$.
It is impossible to discuss here all the good
work that has been done, and is still going on, to attempt to explain
the properties of this state (see {\it e.g.} Chavanis
\cite{Chavanis06a,Chavanis06b,Chavanis06c}, Tsallis {\it et al.}
\cite{Tsallisetal07}, Antoniazzi {\it et al.}
\cite{Antoniazzietal07}, Chavanis {\it et al.} \cite{Chavanisetal08}).
One specific type of numerical simulations,
performed by different groups of authors,
is relevant for this talk.
These have been performed at the
subcritical energy $U=0.69$ with initially all particles located
at the same point ($\theta_i=0$ for all $i$)
and the momenta $p_i$ distributed randomly and uniformly 
in an interval $[-p_{\rm max},p_{\rm max}]$.
The QSS subsequent to these initial conditions has many features
(such as non-Gaussian single-particle velocity distributions) that
have been connected to $q$-statistical mechanics.
Of fairly recent interest is the
sum $Y_i$ of the single-particle momentum $p_i(t)$ sampled at 
regularly spaced times $t=\ell\tau$ along its trajectory,
\beq
Y_i=\sum_{\ell=n_0}^{n_0+N}p_i(\ell\tau).
\label{defyk}
\eeq
The distribution of $Y_i$ in the limit of large $N$ is again controversial
\cite{Pluchinoetal07,Figueiredoetal08}. 
For the specific initial conditions cited above it seems to first approach a
fat-tailed distribution, interpreted by some as a $q$-Gaussian, 
before it finally tends to an ordinary Gaussian.\\

{\it Comments.}
The analogy between (\ref{defy}) and (\ref{defyk}) is obvious. In both cases
the sequence of iterates has long-ranged correlations in the ``time'' variable 
$\ell$ and fills phase space in a lacunary way. 
It is therefore not very surprising that $Y$ and $Y_i$ should have 
{\it non}-Gaussian distributions. 
The $q$-{\it Gaussian\,} shape of these distributions, however, remains
speculative.
%
%
The examples of this talk have shown, on the contrary, that
in the absence of specific arguments sums of correlated variables may have
a wide variety of distributions.
It seems unlikely that haphazard trials
will hit exactly on the $q$-Gaussian.\\

\section{Conclusion}

Universal probability laws occur all around
in physics and mathematics, and
the quest for them is legitimate and interesting.
What lessons can we draw from what precedes?

$\bullet$ It is quite conceivable that new universal distributions may be 
discovered, either by asking new questions about independent variables; or
by asking the traditional questions (sums, maxima,\ldots) about correlated
variables.

$\bullet$ Variables may be correlated in an infinity of ways.
In the end some real-world input is desirable, be it from physics,
finance, or elsewhere. 

$\bullet$ Nothing can beat a central limit theorem.
A good one, however, should give rise to analytic examples
and/or simulation models that reproduce 
the theorem with high numerical precision.

$\bullet$ In the absence of theoretical arguments,
assigning analytic expressions to
numerically obtained curves is a risky undertaking.

Let me end by a quotation \cite{Baars88}:
{\it ``Good theory thrives on reasoned dissent, and} [{\it our views}] {\it may
  change in the face of new evidence and further thought.''}\\

\bigskip

\noindent
{\bf Acknowledgments}
\medskip

The author thanks the organizers of NEXT2008 for this
possibility of presenting his view.
He also thanks Constantino Tsallis for discussions
and correspondence over an extended period of time.



\begin{thebibliography}{99}
\bibitem{footnote1}
The BHP curve is in fact very close to the Gumbel-$k$ distribution with
$k=1.57$ shown in Fig.\,\ref{figgumbel}.
\bibitem{footnote4}
For $\gamma\to 0$ the force $-U'(x)$ is linear and
we recover from (\ref{solqG}) the ordinary Gaussian,
that is, the equilibrium distribution in a harmonic potential.
\bibitem{footnote2}
Ref.\,\cite{Umarovetal08} associates with a function $f(x)$ the
$q$-Fourier transform $\hat{f}_q(\xi)=\int\!\dd x\,f(x)
\big[ 1-(q-1)\,{\rm i}\,\xi xf^{q-1}(x) \big]^{ -1/(q-1) }$.
As an example let us take
$f(x)=( \lambda/x )^{1/(q-1)}$ 
in an interval $[a,b]$ (with $a,b,\lambda>0$)
and $f(x)=0$ zero otherwise. Normalization fixes 
$\lambda$ as a function of $a$ and $b$. 
Then it is easily verified that
$\hat{f}_q(\xi)$ 
is the same for the entire one-parameter family
of intervals defined by $\lambda(a,b)=\lambda_0$.
Hence the $q$-FT is not invertible on the space of probability
distributions.
Other examples may be constructed.
\bibitem{footnote5}
The exact procedure that they followed is described in
Ref.\,\cite{HilhorstSchehr07}.
\end{thebibliography}
\end{document}